\documentclass{pigzpp-whitepaper}
\newcommand{\ResultDate}{2026-07-15}
\newcommand{\ResultCommit}{e7e3286b08fc}

\newcommand{\CliZlibVsPigz}{2.1}
\newcommand{\CliIsalVsPigz}{8.5}

\newcommand{\CliIsalSizePenalty}{9}

\newcommand{\PythonIsalCompetitorMBps}{183}
\newcommand{\PythonZlibMBps}{278}
\newcommand{\PythonIsalMBps}{875}
\newcommand{\PythonZlibVsGzip}{16}
\newcommand{\PythonIsalVsGzip}{50}
\newcommand{\PythonDecompGzipMBps}{192}

\newcommand{\PythonDecompIsalMBps}{525}
\newcommand{\PythonDecompPigzppMBps}{498}
\newcommand{\GoZlibMBps}{294}
\newcommand{\GoIsalMBps}{1087}
\newcommand{\RustZlibMBps}{261}
\newcommand{\RustIsalMBps}{912}

\newcommand{\DockerPgzipMBps}{376}
\newcommand{\DockerZlibMBps}{381}

\newcommand{\DockerZlibVsStdlib}{11}
\newcommand{\DockerIsalVsStdlib}{42}

\newcommand{\DockerStdlibSeconds}{19.6}
\newcommand{\DockerIsalSeconds}{0.5}

\newcommand{\WasmEightMBps}{242}
\newcommand{\WasmEightSpeedup}{5.9}

\newcommand{\PngVsPillow}{11.8}
\newcommand{\PngVsOpenCV}{1.6}
\newcommand{\ZipZlibVsStdlib}{12}
\newcommand{\ZipIsalVsStdlib}{28}

\begin{document}

% ---- Title & abstract (span both columns) ---------------------------------
\twocolumn[{%
  \titlebanner
    {pigzpp}
    {Fast, Parallel, Portable Compression for the Whole Stack}
    {Thamme Gowda \\ Microsoft}
    {https://github.com/thammegowda/pigzpp}
    {An experiment in AI-assisted modernization of foundational software}
  \begin{center}
    {\color{slate}\footnotesize Affiliation for identification only; the views
    expressed here are the author's own.}
  \end{center}
  \begin{minipage}{\textwidth}
    \small
    \noindent\textbf{Abstract.}\enspace
    \texttt{pigz} is a widely deployed parallel gzip utility, but its
    process-global mutable state means that it was not designed as a reentrant,
    directly embeddable library. \textbf{pigzpp} is a from-scratch C++23 rewrite
    that turns the design into a thread-safe library with one accelerated DEFLATE
    core exposed to \textbf{C++, Python, WebAssembly, Go, and Rust}. As
    application-level conveniences built on that same core, it also provides
    multi-entry \textbf{ZIP} archives and a fast \textbf{PNG} codec. Its portable
    zlib-ng backend preserves gzip/zlib's ratio; its Intel ISA-L backend is an
    x86-64-only fast path that produces about 10\% larger output at level~6 on our
    text corpus. In that configuration, zlib-ng reaches
    \best{\CliZlibVsPigz\by} the CLI throughput of \texttt{pigz} and
    \best{\PythonZlibVsGzip\by} that of Python's in-memory \texttt{gzip};
    ISA-L reaches \best{\CliIsalVsPigz\by} and
    \best{\PythonIsalVsGzip\by}, respectively.
    Outputs remain standards-compliant and cross-decode with
    \texttt{gzip}\slash\texttt{pigz}\slash\texttt{unzip}. The result is a
    self-contained, multi-platform compression stack and a case study in
    AI-assisted modernization under automated compatibility tests.

    \vspace{0.5em}

    \vspace{0.75em}
    {\color{rule}\rule{\textwidth}{0.5pt}}\\[0.55em]
    {\centering
    \begin{minipage}{0.82\textwidth}
      \centering
      \includegraphics[width=\linewidth]{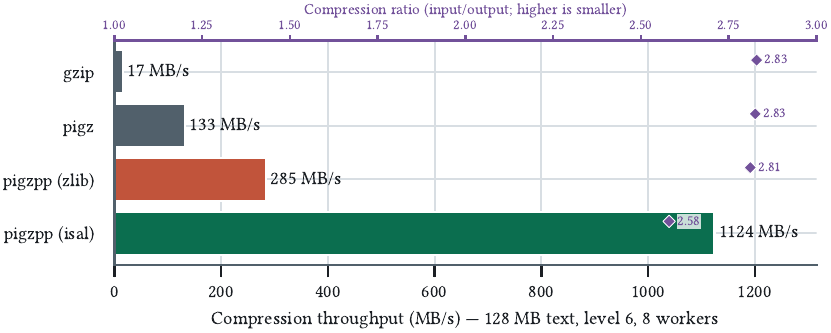}
      \captionof{figure}{\textbf{Headline result.} CLI compression throughput
      on 128\,MB text (level~6, 8 workers). Portable zlib-ng reaches
      \CliZlibVsPigz\by\ \texttt{pigz} at the same ratio; x86-64 ISA-L
      reaches \CliIsalVsPigz\by\ and produces about
      \CliIsalSizePenalty\% more output.}
      \label{fig:cli}
    \end{minipage}
    \par}
  \end{minipage}
  \vspace{0.65em}
  \hypersetup{pageanchor=true}%
}]

% ---- 1. Introduction ------------------------------------------------------
\section{Introduction}
Compression is quiet infrastructure. It sits under package managers, container
builds, log pipelines, scientific datasets, web assets, and archival storage---so
pervasive that most of the software stack pays its cost without ever naming it.
A faster compressor is therefore not a niche optimization; it is a broad,
compounding speedup across everything built on top of it. And speed today means
one thing above all: \emph{parallelism}. Data volumes and CPU core counts have
both grown for years while single-core speed has plateaued, so the central
question for modern compression is how well it turns many cores into throughput.
This paper is about parallel compression meeting modern multi-core hardware.

For three decades, one family of formats has carried the bulk of that load. The
\textbf{DEFLATE} format~\citep{rfc1951}---LZ77~\citep{zivlempel1977} matching
combined with Huffman~\citep{huffman1952} coding---together with
\texttt{gzip}~\citep{rfc1952} and its reference implementation
zlib~\citep{rfc1950}, became the universal default: it is the
\texttt{.gz} file, the common DEFLATE method in \texttt{.zip} archives, the
compressed image-data stream in PNG, the HTTP \texttt{Content-Encoding: gzip}
response, and the OCI/Docker image layer. Its
ubiquity, not its raw ratio, is what makes it irreplaceable---anything that reads
compressed data almost certainly speaks DEFLATE. Yet DEFLATE was designed to run
on a single core, which puts the format the world depends on at odds with the
hardware the world now runs.

\texttt{pigz}\footnote{\texttt{pigz}: \url{https://zlib.net/pigz/}}, written by
Mark Adler, resolved that tension by making
DEFLATE \emph{parallel}: it splits the input into blocks, compresses them across
a thread pool, and stitches the results back into a fully standard gzip stream.
For over a decade \texttt{pigz} has been the go-to tool for saturating a
multi-core machine with backward-compatible gzip.

Yet the shape of software has changed since \texttt{pigz} was written. Today a
compression routine is as likely to be called from Python, a browser, or a Go
service as from a shell, and \texttt{pigz} cannot serve those callers: its logic
is entangled with a process-global \texttt{struct g} of roughly sixty mutable
fields, so it is a program, not a library. As its author put it,
\emph{``pigz is not a library. You would need to adapt the source code in pigz
for use in other applications. That would take some work.''}\footnote{M.~Adler,
Stack Overflow answer, \url{https://stackoverflow.com/a/38946216/1506477}.}
\textbf{pigzpp} asks a simple question---\emph{what would \texttt{pigz} look
like if it were designed today?}---and answers it with a thread-safe library
that keeps the parallel speed, adds hardware-accelerated backends (SIMD zlib-ng
and assembly ISA-L), and exposes one shared core to C++, Python,
Web\-Assembly, Go, and Rust. ZIP and PNG conveniences apply that core to archive
and image workflows rather than introducing new compression engines.

\paragraph{Contributions.} pigzpp provides (1) reentrant, block-parallel gzip in
C++23; (2) portable zlib-ng and x86-64 ISA-L behind one API; (3) bindings for
five ecosystems plus application-level ZIP and PNG conveniences built on the
same DEFLATE core.

\paragraph{Provenance.} pigzpp is also an AI-assisted modernization experiment:
much of the rewrite was performed by coding agents working under human direction,
as documented in two companion essays.\footnote{T.~Gowda, \emph{Making Compression
Faster}, \url{https://gowda.ai/posts/2026/07/fast-compression/}; and \emph{I Let
Two AI Agents Race to Modernize pigz},
\url{https://gowda.ai/posts/2026/03/pigzpp-with-agents/}.} The goal was never to
diminish the original authors' work but to study how well modern agents can
modernize real, load-bearing software.

% ---- 2. Related Work -----------------------------------------------------
\section{Related Work}
Parallel and accelerated DEFLATE has a rich lineage, and pigzpp deliberately
reuses rather than reinvents the hard parts.

\textbf{pigz} pioneered block-parallel gzip and remains the reference
for CLI throughput; pigzpp keeps its parallelization strategy but repackages it
as a reentrant library. \textbf{zlib-ng}\footnote{zlib-ng:
\url{https://github.com/zlib-ng/zlib-ng}} and \textbf{libdeflate}\footnote{libdeflate:
\url{https://github.com/ebiggers/libdeflate}} modernize the DEFLATE inner loops
with SIMD; pigzpp embeds zlib-ng as its ratio-oriented backend rather than
competing with it. Intel \textbf{ISA-L}\footnote{Intel ISA-L:
\url{https://github.com/intel/isa-l}} provides hand-written assembly DEFLATE that
is faster still; on the level-6 text corpus used here, its ratio trade-off means
about 10\% larger output than zlib-ng,
and pigzpp exposes it as the default throughput backend. Per-language parallel
wrappers such as \texttt{klauspost/pgzip} (Go) and \texttt{gzp} (Rust) each
re-solve parallelization inside a single ecosystem; pigzpp instead shares one
accelerated core across five languages, so an improvement lands everywhere at
once.

Newer formats such as \textbf{Zstandard}~\citep{rfc8878},
\textbf{Brotli}~\citep{brotli2019}, and
\textbf{LZ4}\footnote{LZ4: \url{https://github.com/lz4/lz4}} often beat
DEFLATE on the ratio/speed frontier, but they are not wire-compatible with the
enormous installed base of \texttt{gzip}, \texttt{.zip}, PNG, HTTP
\texttt{Content-Encoding: gzip}, and OCI image layers. pigzpp's goal is
orthogonal: make the \emph{format the world already uses} as fast as modern
hardware allows, so it can be dropped in with no changes to existing readers.

% ---- 3. Design ------------------------------------------------------------
\section{Design and Architecture}
\subsection{One accelerated core, many front ends and applications}
pigzpp centers on a single C++23 compression library with no global state.
The CLI and every language binding expose that core directly; the ZIP writer and
PNG encoder are application layers that call it internally. A performance
improvement in the core therefore benefits both low-level APIs and these
higher-level conveniences.

\begin{itemize}
  \item \textbf{Thread-safe by construction} --- configuration is passed
  explicitly; multiple compress/decompress operations run concurrently in one
  process. Threads use C++20 \texttt{std::jthread}; there is no
  \texttt{setjmp}\slash\texttt{longjmp} and no shared mutable state.
  \item \textbf{Selectable backend} --- \texttt{auto} (ISA-L on supported
  x86-64 systems, otherwise zlib-ng), \texttt{zlib} (portable zlib-ng), or
  \texttt{isal}, chosen per call via the API or the \texttt{-{}-engine} CLI flag.
  \item \textbf{Format-compatible} --- the core emits standard gzip, zlib, and
  raw DEFLATE streams; the application helpers wrap those streams in standard
  ZIP or PNG structures. Existing \texttt{gzip}, \texttt{pigz}, \texttt{unzip},
  and image tools can read the corresponding outputs.
\end{itemize}

\subsection{Two DEFLATE backends}
The core dispatches to one of two vendored, statically linked engines:
\textbf{zlib-ng}, a SIMD-optimized drop-in for zlib that maximizes
ratio, and Intel \textbf{ISA-L}, whose hand-written assembly DEFLATE
maximizes throughput. For the highest compression ratio at level~11, pigzpp can
also call Google \textbf{Zopfli}\footnote{Zopfli:
\url{https://github.com/google/zopfli}}. zlib-ng performs runtime CPU
dispatch, so a single portable binary uses the best available SIMD path
(SSE/AVX2/AVX-512 on x86, NEON on ARM) without per-CPU builds.

\paragraph{ISA-L is x86-64 only.} ISA-L's DEFLATE is hand-written x86-64
assembly, so the ISA-L backend is available only on x86-64; on ARM64---and in
WebAssembly, which has no ISA-L build---pigzpp uses zlib-ng. We therefore report
\emph{both} backends throughout: the \texttt{isal} numbers are the x86-64 fast
path, while the \texttt{zlib} (zlib-ng) numbers are the portable path every
platform receives, at gzip/zlib's ratio.
\subsection{Anatomy of a parallel compression call}
To make the pipeline concrete, consider compressing a 128\,MB file at level~6
with 8 threads. pigzpp splits the input into fixed-size blocks (128\,KiB by
default) and hands them to a thread pool. Each worker deflates its block
independently, priming the compressor with the last 32\,KiB of the previous
block as a dictionary so that cross-block back-references---and thus the
compression ratio---are preserved. In parallel, each worker computes the CRC-32
of its raw block; the per-block checksums are then folded into one file checksum
with \texttt{crc32\_combine}, which composes two CRCs in $O(\log n)$ without
rescanning the data. A single writer thread emits the gzip header, streams the
finished blocks strictly in input order, and appends the combined CRC and length
trailer. The result is a byte-stream that any \texttt{gzip} or \texttt{pigz} can
decode, produced while workers remain busy and without a global lock on the hot
path.
\subsection{Language bindings}
The same core is surfaced to five ecosystems:
\begin{itemize}
  \item \textbf{Python} via nanobind\footnote{nanobind:
  \url{https://github.com/wjakob/nanobind}}, shipped as a single
  CPython stable-ABI (\texttt{abi3}) wheel that works on CPython~3.12 and newer.
  \item \textbf{WebAssembly} via Emscripten/Embind, in baseline, SIMD, and
  threaded (\texttt{SharedArrayBuffer}) variants.
  \item \textbf{Go} (cgo) and \textbf{Rust} (FFI) via a stable C ABI.
  \item \textbf{C++} directly as a static or shared library.
\end{itemize}

\subsection{Applications built on DEFLATE: ZIP and PNG}
ZIP and PNG are application conveniences, not additional compression backends.
Both are layered on the same core. The ZIP API mirrors Python's
\texttt{zipfile}: it writes archive metadata, sends each DEFLATE member through
the parallel compressor, and supports STORED entries and Zip64. The PNG codec
filters scanlines and places their zlib-wrapped DEFLATE stream in \texttt{IDAT}
chunks while handling the surrounding PNG structure. It accepts grayscale,
grayscale+alpha, RGB, and RGBA buffers.

% ---- 3. Benchmarks --------------------------------------------------------
\section{Benchmark Results}
Native measurements were made in an Ubuntu~22.04 WSL2 guest whose host-reported
CPU is an Intel Xeon W-2235. Although that processor has six physical cores and
twelve hardware threads, this guest was exposed to only ten logical processors;
Linux reports the allocation as one socket with five cores and two threads per
core. Thus the five- and ten-worker points below represent the guest-visible
core and SMT limits, not the host's full topology.
Unless a caption says otherwise, text results use a 128\,MB multilingual corpus
(English and Chinese Wikipedia), level~6, and eight workers. Every report
benchmark performs one untimed warm-up followed by seven timed samples and uses
the median as its primary result; PNG treats each complete 24-image pass as one
sample. The harnesses do not standardize CPU affinity, turbo state, or frequency
scaling. The archived result set was recorded on \ResultDate\ at commit
\texttt{\ResultCommit}.

\texttt{ratio} is input/output (higher is smaller output). Because ISA-L is
x86-64 only, we report both backends: \texttt{zlib} is the portable zlib-ng path,
whereas \texttt{isal} is the additional x86-64 fast path. Across all plots,
\textbf{green} denotes ISA-L, \textbf{terracotta} denotes zlib-ng, and gray
denotes other implementations. In figures that show two metrics, bars and the
lower axis encode throughput; \textcolor[HTML]{72519A}{purple diamonds, labels,
and the upper axis} encode compression ratio or relative output size.

\subsection{Command line: pigzpp vs.\ gzip and pigz}
Figure~\ref{fig:cli} shows that pigzpp zlib-ng outpaces \texttt{pigz} at the
same ratio, while ISA-L widens the gap; bars show throughput and purple diamonds
show compression ratio.

\begin{figure}[!t]
\centering
\includegraphics[width=0.96\columnwidth]{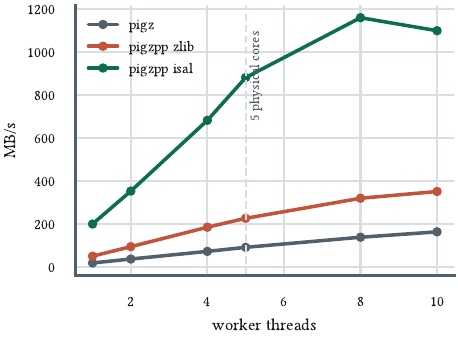}
\caption{Native compression scaling on 128\,MB text (level~6). The dashed line
marks the five cores visible to the WSL2 guest; higher worker counts use its
SMT threads.}
\label{fig:native-scaling}
\end{figure}
\paragraph{Native scaling.} Figure~\ref{fig:native-scaling} shows where the
parallel implementations saturate across the five cores and ten logical CPUs
visible to the WSL2 guest. In the fresh run, ISA-L peaks at eight
workers (1,158\,MB/s) and remains within 6\% at ten, while zlib-ng and
\texttt{pigz} continue scaling through ten workers.

\subsection{Language bindings: pigzpp vs.\ each ecosystem}
\begin{figure}[!t]
\centering
\includegraphics[width=0.96\columnwidth]{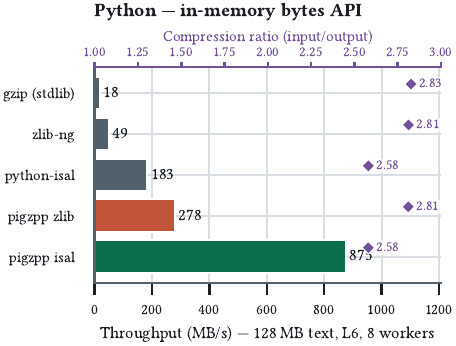}
\caption{Python in-memory bytes API. Both pigzpp backends outperform the tested
Python-native alternatives; ISA-L reaches \PythonIsalMBps\,MB/s.}
\label{fig:lang-python}
\end{figure}

\begin{figure}[!t]
\centering
\includegraphics[width=0.96\columnwidth]{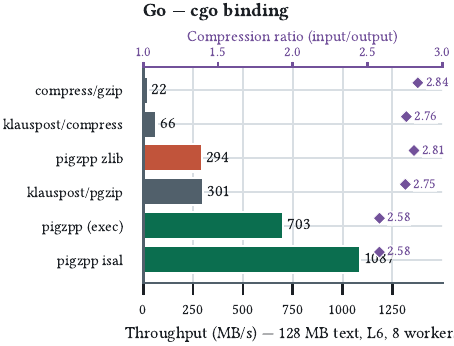}
\caption{Go compression APIs. pigzpp's auto/ISA-L cgo path reaches 1.1\,GB/s;
the subprocess path includes process and pipe overhead.}
\label{fig:lang-go}
\end{figure}

\begin{figure}[!t]
\centering
\includegraphics[width=0.96\columnwidth]{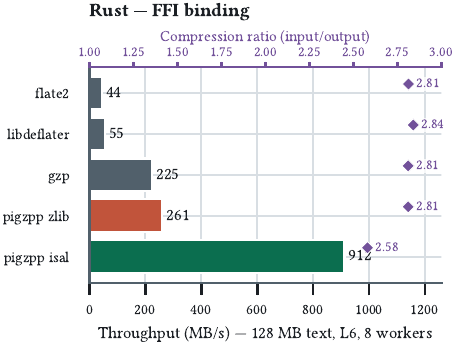}
\caption{Rust compression APIs. pigzpp zlib-ng slightly exceeds parallel
\texttt{gzp}; the ISA-L path reaches \RustIsalMBps\,MB/s.}
\label{fig:lang-rust}
\end{figure}

Figures~\ref{fig:lang-python}--\ref{fig:lang-rust} compare ecosystem-idiomatic
in-memory APIs on the same 128\,MB corpus (level~6, 8 workers). Bars use the
lower throughput axis; purple diamonds use the upper compression-ratio axis.

\paragraph{Python.} The portable pigzpp zlib-ng path reaches
\PythonZlibMBps\,MB/s, while the x86-64 ISA-L path reaches
\best{\PythonIsalMBps\,MB/s}, compared with
\PythonIsalCompetitorMBps\,MB/s for
\texttt{python-isal} in this API-level comparison.

\paragraph{Go.} The cgo paths reach \best{\GoIsalMBps\,MB/s} with ISA-L and
\GoZlibMBps\,MB/s with portable zlib-ng; the subprocess path reaches
703\,MB/s, including process and pipe overhead.

\paragraph{Rust.} zlib-ng reaches \RustZlibMBps\,MB/s versus 225\,MB/s for
\texttt{gzp}; ISA-L reaches \best{\RustIsalMBps\,MB/s}.

\subsection{Docker / OCI image layers}
\begin{figure}[!t]
\centering
\includegraphics[width=0.96\columnwidth]{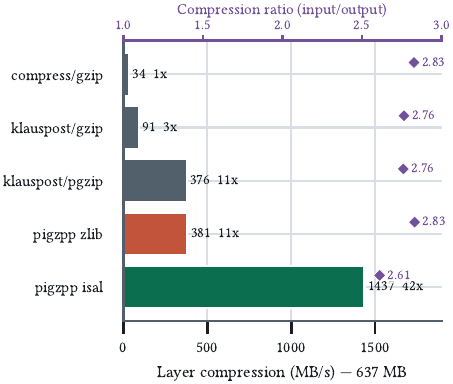}
\caption{Compressing the largest \texttt{python:3.12} image layer (637\,MB); bar
labels give throughput and speed-up over Go's stdlib gzip baseline, while purple
diamonds give ratio. The C-owned pigzpp zlib-ng path reaches
\DockerZlibMBps\,MB/s at ratio 2.83, similar to \texttt{pgzip}'s
\DockerPgzipMBps\,MB/s at ratio 2.76; x86-64 ISA-L reaches
\DockerIsalVsStdlib\by\ the stdlib throughput.}
\label{fig:docker}
\end{figure}

Image layers are gzip-compressed tarballs, and upstream BuildKit's gzip writer
uses Go's single-threaded \texttt{compress/gzip}.\footnote{BuildKit gzip writer:
\url{https://github.com/moby/buildkit/blob/master/util/compression/gzip.go}}
Figure~\ref{fig:docker} isolates that compression stage on a real 637\,MB layer
(the largest layer of \texttt{python:3.12}); it is not an end-to-end
\texttt{docker build} benchmark. Go is built with its default optimizing
compiler and inliner; the C++ core is a Release \texttt{-O3} build. The pigzpp
bars use \texttt{CompressOwnedEngine}, consuming the C-backed output before
release instead of including an additional 225\,MB \texttt{C.GoBytes} copy.

\noindent In this isolated stage, pigzpp compresses the layer
\DockerZlibVsStdlib\by\ faster than the stdlib at the same ratio and
\DockerIsalVsStdlib\by\ faster with ISA-L, reducing measured compression time
from \DockerStdlibSeconds\,s to \DockerIsalSeconds\,s. End-to-end build
speed-up depends on the other build stages.

\subsection{Level and corpus sensitivity}
To test whether the level-6 text result is exceptional, we repeat the native CLI
comparison at levels 1, 6, and 9 on multilingual text and incompressible random
data.

\begin{figure*}[!b]
\centering
\includegraphics[width=0.86\textwidth]{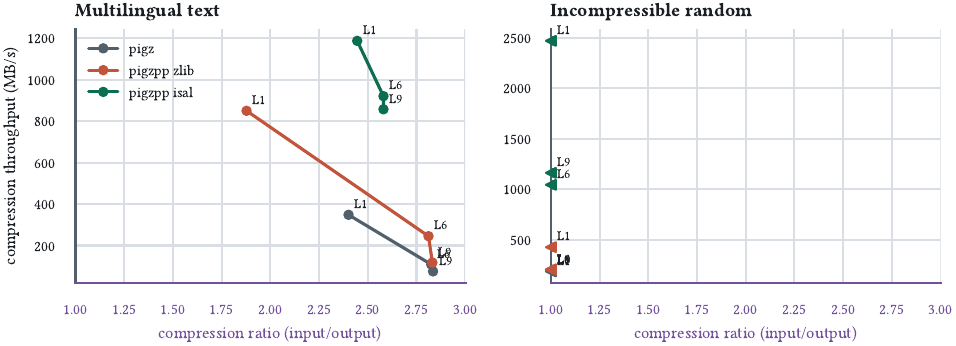}
\caption{Compression throughput versus ratio for levels 1, 6, and 9 on 128\,MB
text and random corpora. Each point is the median of seven timed runs.}
\label{fig:robustness}
\end{figure*}
\noindent Figure~\ref{fig:robustness} reports speed and output ratio together;
in the random panel, left-facing markers at the ratio-1 boundary denote measured
values slightly below~1 due to wrapper overhead.

\subsection{WebAssembly}
\begin{figure}[!t]
\centering
\includegraphics[width=0.96\columnwidth]{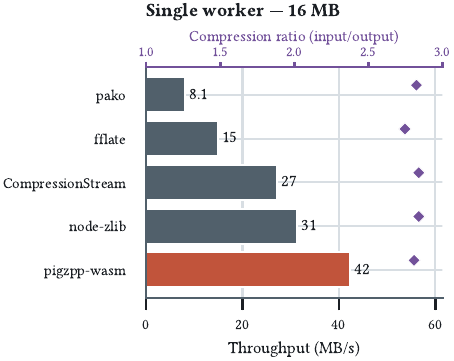}
\caption{Single-worker WebAssembly gzip on 16\,MB text (Node~22). pigzpp-wasm
is fastest among the implementations tested.}
\label{fig:wasm-single}
\end{figure}

\begin{figure}[!t]
\centering
\includegraphics[width=0.96\columnwidth]{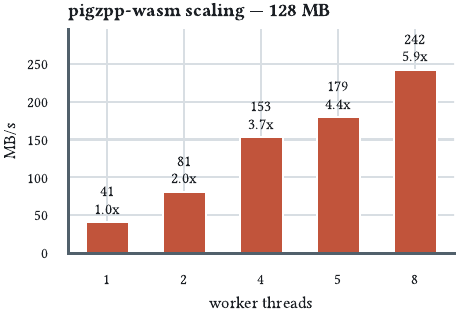}
\caption{Threaded pigzpp-wasm on 128\,MB text. Eight workers reach
\WasmEightMBps\,MB/s in the WSL2 guest allocation of 5 cores/10 logical CPUs.}
\label{fig:wasm-scaling}
\end{figure}

pigzpp ships as WebAssembly (zlib-ng + 128-bit SIMD; ISA-L is x86-only and thus
unavailable in WASM). Figure~\ref{fig:wasm-single} compares single-worker
engines; Figure~\ref{fig:wasm-scaling} measures the threaded build.

\noindent Among the implementations tested, pigzpp-wasm is fastest even with one
worker, ahead of Node zlib, \texttt{CompressionStream}, \texttt{fflate}, and
\texttt{pako}.

\noindent The threaded build reaches 3.72\by\ speed-up with four workers (93\%
worker efficiency); at eight workers it uses SMT and reaches
\best{\WasmEightMBps\,MB/s}, or \WasmEightSpeedup\by\ speed-up (74\% worker
efficiency).

\subsection{PNG application: image encoding}
\begin{figure}[!t]
\centering
\includegraphics[width=0.96\columnwidth]{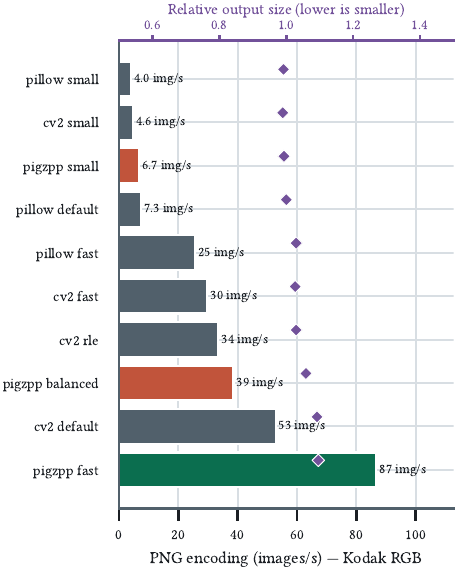}
\caption{Application-level PNG encoding on the 24-image Kodak set
(768$\times$512), RGB. Bars give throughput; purple diamonds give output size
relative to Pillow's default. With ISA-L powering its default \texttt{fast}
preset, pigzpp encodes at
\best{\PngVsPillow\by} Pillow's speed and \best{\PngVsOpenCV\by} OpenCV's, at
comparable size.}
\label{fig:png}
\end{figure}

PNG stores filtered scanlines in a zlib-wrapped DEFLATE stream. We compare the
\texttt{pigzpp.png} convenience codec, which applies the shared core to this
workflow, with Pillow and OpenCV on the 24-image Kodak true-color
set\footnote{Kodak Lossless True Color Image Suite:
\url{https://r0k.us/graphics/kodak/}} (768$\times$512), median of seven complete
passes after one warm-up, with round-trips verified.
\texttt{size} is output relative to Pillow's default (lower is smaller); see
Figure~\ref{fig:png}. pigzpp performs the PNG line filtering itself and then
DEFLATE-compresses the filtered scanlines, so it can route PNG through either
backend. The default \texttt{fast} preset feeds the filtered scanlines to
\textbf{ISA-L} (x86; zlib-ng elsewhere)\footnote{ISA-L is x86-64 only; on ARM64
and in WebAssembly the \texttt{fast} preset uses zlib-ng. A dedicated ARM64
measurement remains future work.}, which is
$\sim$1.6\by\ faster than zlib-ng's \texttt{Z\_RLE} path at equal-or-better size;
the higher-ratio \texttt{balanced} and \texttt{small} presets keep zlib-ng's
strategy-aware modes.

\noindent pigzpp's \texttt{fast} preset is the fastest PNG encoder we
benchmarked on true-color images --- \best{\PngVsOpenCV\by} faster than OpenCV
(itself a hand-optimized C++ encoder) and \best{\PngVsPillow\by} faster than
Pillow's default --- at a comparable size, after routing its filtered scanlines
through ISA-L.

\subsection{ZIP application: archives (Python)}
\begin{figure}[!t]
\centering
\includegraphics[width=0.96\columnwidth]{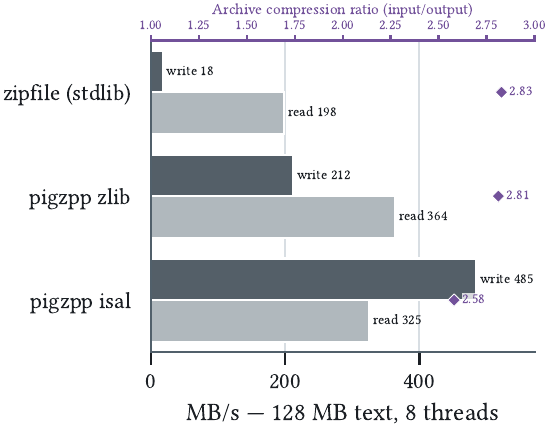}
\caption{Application-level ZIP archives in Python (128\,MB text, 8 threads).
Write/read bars use the lower throughput axis; purple diamonds give archive
ratio. pigzpp applies its parallel DEFLATE core to each member, writing
\ZipZlibVsStdlib--\ZipIsalVsStdlib\by\ faster than the standard library.}
\label{fig:zip}
\end{figure}

ZIP is a container format; here pigzpp provides a convenience API that applies
its parallel DEFLATE core independently to each archive member.
\texttt{pigzpp.ZipFile} vs.\ the standard library's \texttt{zipfile}, both
writing real DEFLATE archives and reading them back (128\,MB text, 8 threads);
results are in Figure~\ref{fig:zip}.

\noindent pigzpp parallelizes each member, writing \ZipZlibVsStdlib\by\ faster
than \texttt{zipfile} at the same ratio (\texttt{zlib}) and up to
\ZipIsalVsStdlib\by\ faster with ISA-L, while reading up to 1.8\by\ faster.

\subsection{Decompression}
The same report run also measures decompression of a common level-6 gzip stream
through native CLIs and Python in-memory APIs.

\begin{figure}[!t]
\centering
\includegraphics[width=0.96\columnwidth]{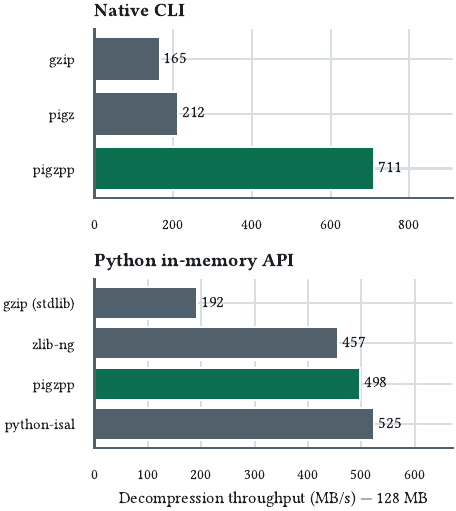}
\caption{Decompression throughput on 128\,MB text. Native tools decode the same
stdlib gzip stream; the Python panel compares in-memory APIs.}
\label{fig:decompression}
\end{figure}
\noindent Figure~\ref{fig:decompression} separates decompression performance from
the encoder-specific speed/ratio trade-offs shown above. Native pigzpp reaches
711\,MB/s, versus 212\,MB/s for \texttt{pigz} and 166\,MB/s for \texttt{gzip};
the Python in-memory results range from \PythonDecompGzipMBps\,MB/s for the
standard library to \PythonDecompIsalMBps\,MB/s for \texttt{python-isal}, with
pigzpp at \PythonDecompPigzppMBps\,MB/s. Python decompression is
single-threaded: its speed comes from optimized inflate, and the binding writes
directly into the result \texttt{bytes} allocation to avoid a full-output copy.

% ---- 4. Use cases ---------------------------------------------------------
\section{Where pigzpp Helps}
\begin{itemize}
  \item \textbf{Command-line and pipelines} --- a drop-in \texttt{gzip}\slash\texttt{pigz}
  replacement for build systems, backups, and log rotation.
  \item \textbf{Container builds} --- faster OCI/Docker layer compression through
  cgo; overall build speed-up depends on non-compression stages.
  \item \textbf{Data and ML} --- high-throughput \texttt{gzip}\slash\texttt{zlib}
  from Python without leaving the process, plus a PNG convenience layer that
  applies the same DEFLATE core to image pipelines.
  \item \textbf{Archives} --- a \texttt{zipfile} convenience layer that applies
  the core per member and produces standard, interoperable ZIPs.
  \item \textbf{Browser and edge} --- WebAssembly builds that outperform native
  \texttt{CompressionStream} and popular JS libraries for both DEFLATE and ZIP.
  \item \textbf{Services in Go and Rust} --- one accelerated core behind idiomatic
  bindings, replacing per-language parallel-gzip libraries.
\end{itemize}

% ---- 5. Packaging ---------------------------------------------------------
\section{Packaging and Portability}
pigzpp is engineered to ship as self-contained artifacts. Releases include a
native CLI archive and a Python \texttt{abi3} wheel for each of
\{Linux, macOS, Windows\}~$\times$~\{x86-64, ARM64\}, plus three WebAssembly
variants. A single stable-ABI wheel serves CPython~3.12 and every later release,
subject to that ABI's platform support, removing the usual per-version wheel
matrix.

Crucially, every artifact \textbf{embeds its compression engines statically}:
zlib-ng, ISA-L (where its assembly is supported), Zopfli, and the binding runtime
are linked in, so no artifact requires those libraries to be installed at run
time. Linux CLI binaries are fully static; Windows binaries use the static MSVC
runtime; Linux wheels additionally embed the GNU C++/GCC runtimes. Continuous
integration inspects every final archive and repaired wheel and \emph{rejects}
any dynamic dependency on zlib-ng or ISA-L, guaranteeing the self-contained
property rather than merely assuming it.

% ---- 6. Reproducibility ---------------------------------------------------
\section{Reproducibility}
Each reported value originates in a harness under \texttt{benchmarks/}: core,
Python, PNG, Go/Docker, Rust, or WASM. Core, language-binding, ZIP, and WASM
experiments share corpora generated by \path{benchmarks/core/gen_data.py};
Docker uses a real image layer, and PNG uses the Kodak image set. Representative
commands are:

\begin{center}
\begin{minipage}{0.96\columnwidth}
\ttfamily\scriptsize\color{ink}
make bench-bin\ \ \# CLI: gzip vs pigz vs pigzpp\\
make bench-py\ \ \ \# Python: gzip/zlib-ng/isal/pigzpp\\
make bench-png\ \ \# PNG: pigzpp vs Pillow / OpenCV
\end{minipage}
\end{center}
\noindent Absolute throughput varies with CPU, scheduler, frequency state, and
library versions. The TSV files under \texttt{report/plots/data/} are generated
from archived raw JSON under \texttt{benchmarks/results/}; the JSON records every
timed sample, exact command, commit state, machine/runtime metadata, corpus sizes
and SHA-256 hashes, output sizes, and ratios. \texttt{make plots} regenerates all
figures, and \texttt{benchmarks/report/validate\_report\_data.py} rejects missing
samples or TSVs that disagree with the raw results.

% ---- Limitations ---------------------------------------------------------
\section{Limitations and Threats to Validity}
Native results come from one x86-64 WSL2 environment with a 10-vCPU guest
allocation; no native ARM64 throughput is reported, and ISA-L results do not
generalize beyond x86-64. Most compression
comparisons use one 128\,MB multilingual-text corpus at level~6, so speed and
ratio may differ for source code, binary data, incompressible input, and other
levels. The language panels compare ecosystem-idiomatic APIs rather than
isolating binding overhead. No common CPU affinity, turbo, or frequency-control
policy is enforced. Accordingly, ``fastest'' means fastest
among the implementations, inputs, and configurations tested here; it is not a
universal ranking.

% ---- Lessons Learned -----------------------------------------------------
\section{Lessons Learned}
Two lessons stand out from rebuilding a mature, load-bearing C program with AI
coding agents.

\textbf{AI agents excel when feedback is tight.} Progress was fastest where
correctness was mechanically checkable: lossless round-trips, cross-decoding with
\texttt{gzip} and \texttt{unzip}, a large test suite, and repeatable benchmarks.
Agents propose changes quickly; automated oracles are what make accepting those
changes safe.

\textbf{Global state is the real portability barrier.} The hardest part of the
rewrite was not performance but untangling \texttt{pigz}'s process-global
\texttt{struct g}. Removing it---threading configuration explicitly and giving
each operation its own state---is what turned a program into a library and
unlocked every binding. Performance largely took care of itself once the
vendored engines were in place.

% ---- Conclusion ----------------------------------------------------------
\section{Conclusion}
pigzpp shows that a decades-old, ubiquitous format need not be slow. By pairing
\texttt{pigz}'s parallel design with SIMD and assembly DEFLATE backends and
exposing one thread-safe core to five languages, its portable zlib-ng path reaches
up to \best{\CliZlibVsPigz\by} the CLI throughput of \texttt{pigz} and
\best{\PythonZlibVsGzip\by} that of Python's in-memory \texttt{gzip}; its
x86-64-only ISA-L path reaches \best{\CliIsalVsPigz\by} and
\best{\PythonIsalVsGzip\by}, respectively. Its output remains
standards-compliant and interoperable as gzip and through the ZIP and PNG
application layers built on the same DEFLATE core. Release artifacts
embed pigzpp's compression engines and relevant C++ runtime dependencies, while
using required platform system libraries. The benchmark harnesses and exact plot
inputs are included so the evaluation can be inspected and extended.

% ---- 7. Credits -----------------------------------------------------------
\section{Acknowledgments and Credits}
pigzpp stands entirely on the shoulders of prior work, and this section is the
most important one.

\begin{itemize}
  \item \textbf{Mark Adler} --- co-creator of zlib, gzip, and the DEFLATE
  format~\citep{rfc1950,rfc1952,rfc1951}, and author of \texttt{pigz}. pigzpp is a
  rewrite of
  his design and exists only because of the decades of work he invested in
  building and maintaining these foundations. This is an \emph{altered} version;
  it is not the original \texttt{pigz} and must not be mistaken for it.

  \item \textbf{Marcin Junczys-Dowmunt} --- for valuable discussion,
  and in particular for suggesting that combining \texttt{pigz} with Intel ISA-L
  could yield large speedups on Intel x86 hardware, an idea that became pigzpp's
  default throughput backend.

  \item \textbf{zlib-ng contributors} --- the SIMD-optimized,
  runtime-dispatched DEFLATE engine that gives pigzpp its portable acceleration.
  \item \textbf{Intel ISA-L} --- the hand-tuned assembly DEFLATE that
  powers pigzpp's fastest path.
  \item \textbf{Google Zopfli} --- optimal DEFLATE for the
  highest-ratio level.

\end{itemize}

\noindent The design rationale, the AI-assisted process, and the deeper
performance analysis are described in the two companion essays footnoted in the
introduction. pigzpp is released under the zlib license, the same license as the
original \texttt{pigz}.

% ---- References (published works only; software/blogs are footnoted inline) --
\bibliographystyle{acl_natbib}
\bibliography{references}

\end{document}